\documentclass[%
 amsmath,amssymb,
 aps,
pra,
floatfix,
]{revtex4-2}

\usepackage[dvipdfmx]{graphicx}
\usepackage{dcolumn}
\usepackage{bm}
\usepackage{algorithm}
\usepackage{algorithmic}
\usepackage{color}
\usepackage{empheq}

\begin{document}

\preprint{APS/123-QED}

\title{Dynamics-Based Intrinsic Signal Model for High-Dimensional, Small-Sample Data}

\author{Yoh-ichi Mototake}
 \email{y.mototake@r.hit-u.ac.jp}
 \homepage{https://researchmap.jp/mototake/} 
\affiliation{%
 Graduate School of Social Data Science, Hitotsubashi University, 2-1 Naka, Kunitachi-shi, Tokyo 186-8601, Japan}%

\author{Y-h. Taguchi} \email{tag@granular.com}
 \homepage{https://researchmap.jp/Yh\_Taguchi/} 
\affiliation{
 Department of Physics, Chuo University, 1-13-27 Kasuga, Bunkyō-ku, Tokyo 112-8551 Japan}%

\date{\today}

\begin{abstract}
Signal extraction is difficult when the number of variables $N$ is much larger than the number of observations $M$. We address this problem under the working hypothesis that an empirical dataset consists of states sampled from underlying multivariate dynamics. Instead of treating the $M$ observations as points in an $N$-dimensional variable space, we treat the $N$ variables as points in an $M$-dimensional sample-coordinate space, interpreted as an effective time-delay coordinate space of the latent dynamics. This representation allows a large $N$ to provide many points for estimating the variable distribution even when $M$ is small. Transposed singular value decomposition (SVD) and the unsupervised feature-selection method of Taguchi are used to extract variable-side deviations from an estimated Gaussian background as signal candidates. As $M$ is reduced, the effective separation between sampled states increases; contributions from finite-correlation components are expected to decay, whereas sufficiently long-correlation components can persist toward the small-sample limit. We define these persistent components as intrinsic signals and estimate them by extrapolation toward $M=0$. We first tested the method on high-dimensional, small-sample data explicitly generated by a randomised coupling strength globally coupled map (RCS-GCM), for which the long- and short-correlation components were known. The extracted intrinsic signals corresponded to the known long-correlation variables. We then applied the method to The Cancer Genome Atlas (TCGA) pan-kidney gene-expression data, which are not ordinarily treated as data explicitly generated by a dynamical system. Using an SVD component associated with the pathologic-M category, variable-side signal components were extracted from the 20,531-dimensional data even under small-sample conditions. These results indicate that the proposed dynamics-based intrinsic-signal analysis is also effective for high-dimensional, small-sample empirical data without explicit temporal information.
\end{abstract}

\maketitle

\section{Introduction}
\label{sec_intro}

Signal extraction is generally formulated as a variable-selection problem that depends on either an externally specified objective or a criterion chosen by the analyst~\cite{brunton2016discovering,uemura2015variable}. In supervised learning, variables are regarded as signals when they are useful for predicting a target variable. Unsupervised methods avoid an explicit target but still employ criteria specified in advance. For example, principal component analysis identifies directions of maximal variance~\cite{Pearson1901} and commonly selects the number of retained components using explained-variance criteria~\cite{Jolliffe2002,Jackson1991}, whereas clustering assumes that distance represents similarity~\cite{Luxburg2007}. Recent foundation models~\cite{bommasani2021opportunities} show that reusable representations can be learned without defining a separate supervised objective for every downstream task~\cite{achiam2023gpt}. However, they do not determine which variables in a given dataset should be regarded as signals from the data-generating process itself; rather, the relevance of variables or representations is ultimately evaluated operationally through their usefulness for downstream tasks. In conventional formulations, regardless of the criterion used, identifying signal variables requires estimating the corresponding structure of the data distribution in variable space from a finite set of observations. In high-dimensional, small-sample settings, however, such estimation can become unreliable. For example, conventional PCA eigenvalue estimators can be directly affected by high-dimensional noise and become inconsistent~\cite{YataAoshima2013}. High-dimension, low-sample-size asymptotic theory has therefore been developed to address such difficulties~\cite{AoshimaEtAl2018}. Consequently, when the number of available observations is limited relative to the dimensionality of the data, reliable estimation of the structure required for signal identification becomes difficult.\par

In this study, we define signals on the basis of the dynamics that generate the data, rather than solely through an externally specified objective or an analyst-chosen criterion. We adopt the working hypothesis that any empirical dataset can be regarded as a collection of states sampled from underlying multivariate dynamics. Here, the dataset need not be explicitly observed as a time series. For example, natural images of animals may reflect long-timescale evolutionary dynamics as well as shorter-timescale processes associated with the state of the animal and image acquisition. Likewise, apparently static ecological snapshots can retain signatures of underlying ecosystem dynamics, as reflected in spatial indicators such as spatial correlations and spectral properties~\cite{Kefi2014}. Under this working hypothesis, structures appearing in a dataset can be regarded as consequences of latent data-generating dynamics, providing a basis for defining signals intrinsically from those dynamics.\par

\begin{figure}[tb]
    \centering
    \includegraphics[width=1.0\linewidth]{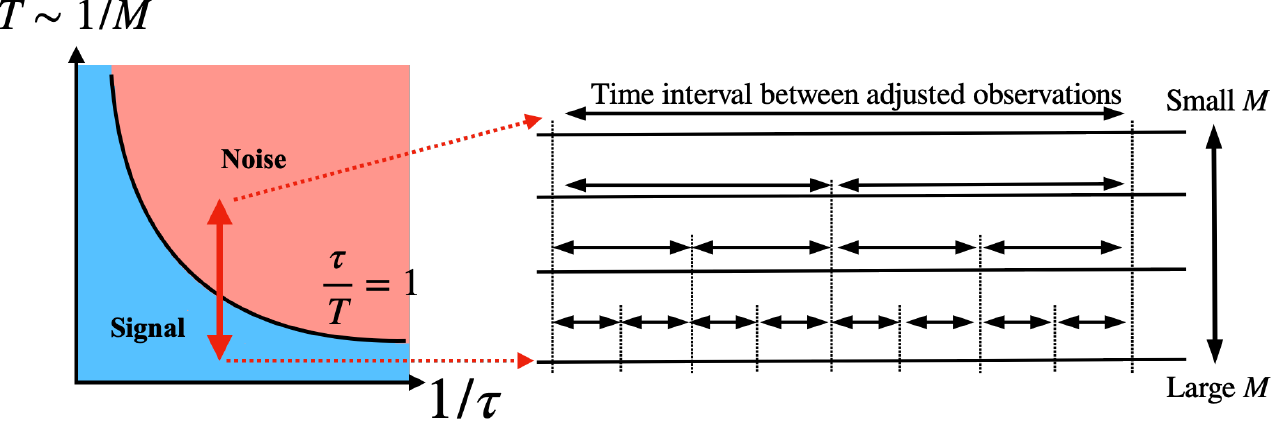}
    \caption{Relationship between the subsample size $M_{\rm res}$ and the effective interval between retained observations. Reducing $M_{\rm res}$ increases the effective delay represented in the time-delay coordinate space. Contributions with finite correlation lengths are expected to decay as the delay increases, whereas a non-vanishing limiting contribution indicates a component whose correlation length is effectively infinite relative to the observation scale.}
    \label{fig:time}
\end{figure}

Under this working hypothesis, we propose that efficient signal extraction from high-dimensional, small-sample data can be achieved by reversing the conventional roles of dimensionality and sample size. Conventionally, the data distribution is represented by observations as points in a high-dimensional variable space. In the present framework, each variable is instead represented by its values across the sampled states and treated as a point in a sample-coordinate space whose dimension is determined by the number of observations. From the dynamical viewpoint, this space is interpreted as an effective time-delay coordinate space of the latent data-generating dynamics~\cite{takens2006detecting, sauer1991embedology}. Thus, a large number of variables provides many points for estimating the variable distribution, whereas the number of observations determines the dimension of the coordinate space. Even when only a small number of observations are available, high dimensionality can therefore provide many points for estimating the variable distribution. 
Within this dynamical interpretation, the detectability of a signal depends on the effective separation between sampled states relative to the correlation time of the underlying dynamics. Reducing the number of retained observations increases the effective separation between sampled states. Components with correlation times short relative to this separation are therefore expected to disappear as the sample size is reduced, whereas components with sufficiently long correlation times can remain detectable toward the small-sample limit as shown in Figure \ref{fig:time}. In this study, such persistent components are defined as intrinsic signals. To estimate these intrinsic signals from the variable distribution in the sample-coordinate space, we use the transposed singular value decomposition (SVD)~\cite{hastie2009elements} feature-selection method of Taguchi~\cite{Taguchi2024}, which has been applied to high-dimensional biomedical datasets with small sample sizes~\cite{Taguchi2024}. The proposed framework thereby aims to extract signals from high-dimensional, small-sample data on the basis of the data-generating dynamics, rather than solely through an externally specified objective or an analyst-chosen criterion. \par

We first evaluated the effectiveness of the proposed dynamics-based intrinsic-signal analysis using a dataset generated by a randomised coupling strength globally coupled map (RCS-GCM) dynamical system, for which the long- and short-correlation components were known. We then applied the same analysis to The Cancer Genome Atlas (TCGA) pan-kidney gene-expression data, which are not ordinarily treated as observations explicitly generated by a known dynamical system, to evaluate the effectiveness of the dynamics-based intrinsic-signal analysis for a high-dimensional empirical dataset without explicit temporal information.

\section{Dynamics-based intrinsic signal model}
\label{sec_method}

\subsection{Time-delay representation of high-dimensional data}
Consider a dataset
\begin{equation}
X=(x_{ij})\in\mathbb{R}^{N\times M},
\end{equation}
where $i=1,\ldots,N$ indexes variables and $j=1,\ldots,M$ indexes observations. In the conventional representation, the dataset consists of $M$ points in the $N$-dimensional variable space. This study considers the point set
\begin{equation}
\mathcal{D}_M=\left\{\mathbf{x}_i=(x_{i1},\ldots,x_{iM})\in\mathbb{R}^{M}\right\}_{i=1}^{N}.
\end{equation}
Each variable is represented as one point in the $M$-dimensional sample space, which is regarded as an effective time-delay coordinate space of the underlying data-generating dynamics. Therefore, the original high-dimensional, small-sample dataset is reinterpreted as $N$ variable points in an $M$-dimensional space.\par
When $N\gg M$, the number of variable points is large relative to the dimension of the variables: $N$ determines the number of points used to estimate the variable distribution, whereas $M$ determines the dimension of the space. These variable points are used to estimate the distribution even when the original number of observations is small. This change in the roles of $N$ and $M$ is the basis of signal extraction in the proposed framework.\par

\subsection{Dynamical structures and intrinsic signals}
\label{signal_model}
\begin{figure}
    \centering
    \includegraphics[width=1.0\linewidth]{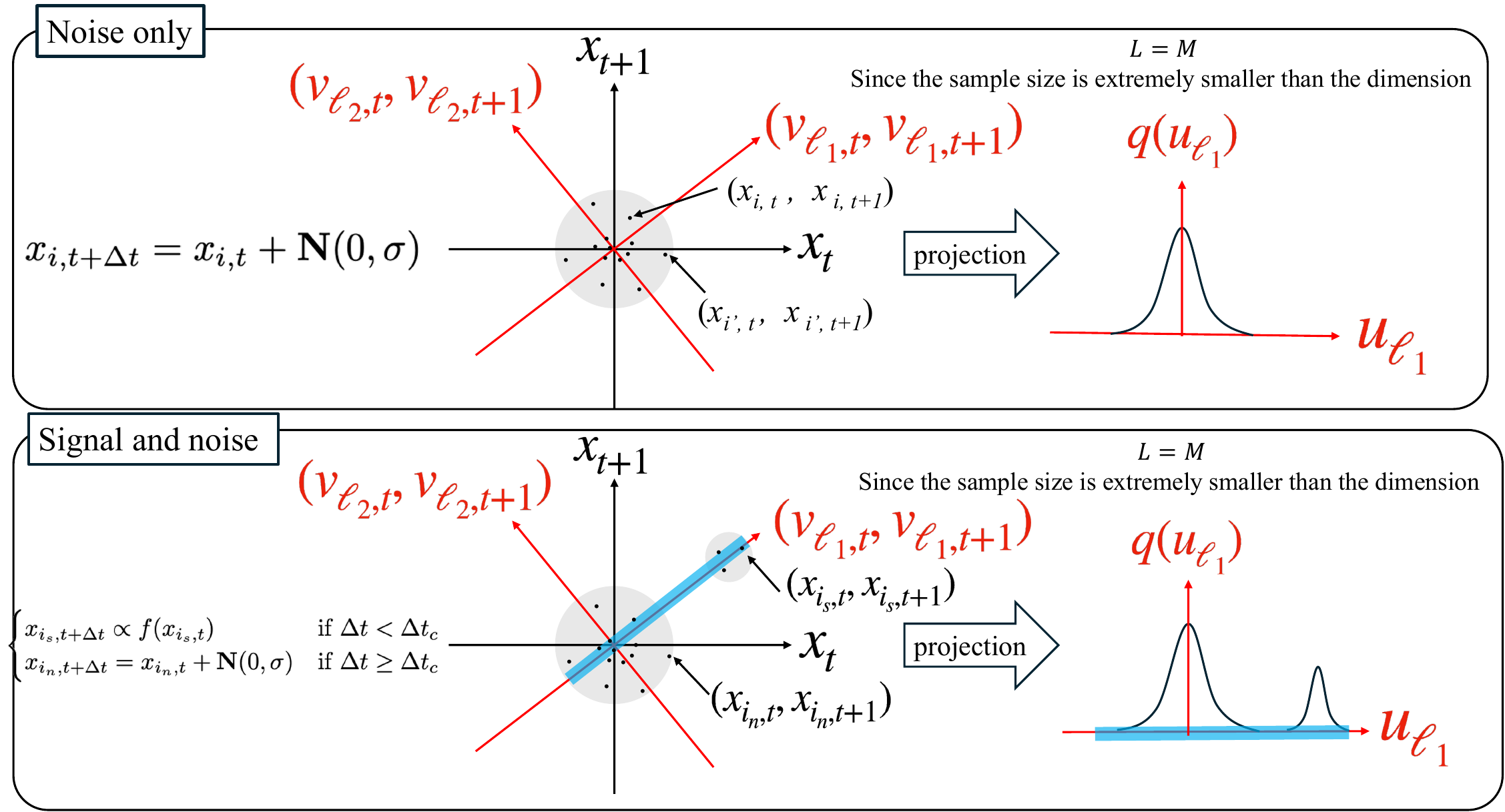}
    \caption{{\bf Upper panel}: Distribution of the variable points when all variables follow random-walk dynamics. For the two sample coordinates $t$ and $t+1$, the pairs $\{(x_{i,t},x_{i,t+1})\}_{i=1}^{N}$ are regarded as $N$ realisations of the two-dimensional random variable $(x_{(\cdot,t)},x_{(\cdot,t+1)})$, with the variable index $i$ serving as the sampling index. The horizontal and vertical axes are denoted by $x_t$ and $x_{t+1}$, respectively. Red arrows denote the SVD directions $(v_{\ell,t},v_{\ell,t+1})$ in this two-dimensional section. The right panel shows the empirical distribution $q(u_{\ell_1})$ of the projected coefficients $\{u_{\ell_1 i}\}_{i=1}^{N}$.{\bf Lower panel}: Corresponding distribution when most variables follow random-walk dynamics and a small number follow structured deterministic dynamics. The structured variables appear as deviations from the random background in the sample-coordinate space and produce non-Gaussian deviations in the projected distribution $q(u_{\ell_1})$.}
    \label{fig:randomwalk}
\end{figure}

Within this framework, signal candidates are distinguished from effectively random components through the variable distribution in the time-delay coordinate space as shown in Figure \ref{fig:randomwalk}. A component whose correlation is unresolved at the observation scale is represented by the random-walk model
\begin{eqnarray}
x_{i,t+\Delta t}=x_{i,t}+\mathbf{N}(0,\sigma).
\end{eqnarray}
where, for readability, we use the notation $x_{i,j}:=x_{ij}$. 
Such components do not form a systematic low-dimensional distribution in the time-delay coordinate space. When sufficiently many variables are available, their projections are approximated by a Gaussian background.\par
Next, consider a system containing components with resolved and unresolved correlations,
\begin{eqnarray}
   \begin{cases}
    x_{i_s,t+\Delta t} \propto f(x_{i_s,t}) & \text{if $\Delta t < \Delta t_c$,} \\
    x_{i_n,t+\Delta t} = x_{i_n,t} + \mathbf{N}(0,\sigma) & \text{if $\Delta t \geq \Delta t_c$,}
    \end{cases}
\end{eqnarray}
where $\Delta t_c$ is the correlation length and $f$ is an arbitrary bounded function. The dependence represented by $f$ produces a non-Gaussian contribution to the variable distribution. For example, when $f(x)=x$, the corresponding variable points are concentrated around $x_{i_s,t+\Delta t}=x_{i_s,t}$, whereas components represented by the random-walk model form a diffuse background. Variable components associated with such deviations are treated as signal candidates. Intrinsic signals are then identified as those signal components that persist toward the small-sample limit.\par

\subsection{Signal estimation by transposed SVD}
The unsupervised feature-selection approach proposed by Taguchi~\cite{Taguchi2024} is used to analyse the distribution of variable points in the effective time-delay coordinate space. Let $x_{ij}\in\mathbb{R}^{N\times M}$ denote the $j$th observation of the $i$th variable. Before decomposition, the $N$-dimensional features of each observation are centred so that their mean across variables is zero. This centring is analogous to the mean-subtraction step of layer normalisation~\cite{ba2016layer}, although no variance scaling is applied.\par
SVD is applied as
\begin{equation}
    x_{ij}=\sum_\ell \lambda_\ell u_{\ell i}v_{\ell j},
\label{eq_svd}
\end{equation}
where $u_{\ell i}$ gives the coordinate of variable $i$ along the $\ell$th principal direction in the $M$-dimensional sample space, and $v_{\ell j}$ describes the corresponding structure over the observations. In conventional SVD,
\begin{equation}
    X=U\Sigma V^t,
\label{eq_normal_svd}
\end{equation}
whereas the present interpretation corresponds to
\begin{equation}
    X^t=V\Sigma U^t.
\end{equation}
Therefore, the analysis estimates directions in the $M$-dimensional sample space from the $N$ variable points $\{(x_{i1},\ldots,x_{iM})\}_{i=1}^{N}$. When $N\gg M$, the number of points used to estimate this distribution is large relative to the dimension of the space.\par
$u_{\ell i}$ represents the variable structure and $v_{\ell j}$ represents the sample structure. Because SVD occurs in the function space, it is easily understood when considered as a Fourier transform. Namely, the sine-wave-like basis functions correspond to $v_{\ell j}$ and the Fourier coefficients (spectra) correspond to $\lambda_\ell u_{\ell i}$. Therefore, a spectrum, $\lambda_\ell u_{\ell i}$, that does not correspond to an interpretable basis function, $v_{\ell j}$, cannot be interpreted.\par
As the sample space is interpreted as a time-delay coordinate space, $u_{\ell i}$ can be written as
\begin{eqnarray}
\label{svd_uli}
    u_{\ell i} &=& \frac{1}{\lambda_\ell}\sum_{j=1}^M x_{ij}v_{\ell j},\label{eq_uli}\\
    &=& \frac{1}{\lambda_\ell}\sum_{t=1}^M x_{it}v_{\ell t}.
\end{eqnarray}
For each SVD component $\ell$, the coefficients
$\{u_{\ell i}\}_{i=1}^{N}$ are regarded as $N$ realisations of a
random variable $u_\ell$, with the variable index $i$ serving as the
sampling index. Its empirical distribution is
\begin{equation}
q(u_\ell)=N^{-1}\sum_{i=1}^{N}\delta(u_\ell-u_{\ell i}).
\end{equation}
Under the null model of effectively random dynamics, this distribution
is assumed to have a Gaussian background. Variables associated with
dynamical structures appear as deviations from this background.
Let
\(
A=\{1,\ldots,N\}
\)
denote the set of all variable indices. Denoting the background-variable
set by $I\subset A$, its complement is defined as
\(
I'=A\setminus I.
\)
The corresponding empirical distributions are
\begin{eqnarray}
q_{\ell,I}(u_\ell)
&=&
\frac{1}{|I|}
\sum_{i\in I}\delta(u_\ell-u_{\ell i})
\sim \mathcal{N}(0,\sigma_\ell^2),\\
q_{\ell,I'}(u_\ell)
&=&
\frac{1}{|I'|}
\sum_{i\in I'}\delta(u_\ell-u_{\ell i})
\not\sim \mathcal{N}(0,\sigma_\ell^2).
\end{eqnarray}
The $P$-value for variable $i$ is evaluated using
\begin{equation}
    P_i=P_{\chi^2}\left[>\left(\frac{u_{\ell i}}{\sigma_\ell}\right)^2\right],
    \label{eq:Pi}
\end{equation}
where $P_{\chi^2}[>x]$ is the upper-tail cumulative $\chi^2$ probability and $\sigma_\ell$ is the standard deviation (SD) of the Gaussian background. Variables with sufficiently small corrected $P$-values are extracted as signal candidates. This variable-side distributional test remains feasible for small $M$ because it is based on the $N$ variable points.\par

\subsubsection{Multiple-comparison correction and persistent-component analysis}
\label{intrinsic_signal_extraction}
As the test is applied to many variables, the Benjamini--Hochberg (BH) method~\cite{benjamini1995controlling} is used to control the false discovery rate. The background SD $\sigma_\ell$ is estimated using the method of Taguchi and Turki~\cite{taguchi2022adapted,Taguchi2024}, which selects the value that makes the non-outlier portion of the $1-P_i$ histogram as uniform as possible. A naive SD estimate based on all variables can be inflated by non-Gaussian outliers, thereby reducing their statistical significance. We therefore estimate $\sigma_\ell$ from the putative Gaussian background after excluding candidate outliers.\par
For each subsample size $M_{\rm res}$, subsets of the observations are drawn repeatedly, transposed SVD is applied, and the variables selected in more than $p\%$ of the trials are counted, denoted $n_{>p\%}(M_{\rm res})$. The dependence of $n_{>p\%}$ on $M_{\rm res}$ is then fitted with a model containing a constant term $n_0$ and a decaying finite-correlation term; the fitted $n_0$ estimates the component that persists under extrapolation to $M_{\rm res}=0$.\par
Thus, the signal component of the variable structure $u_{\ell i}$ is estimated by the Taguchi method. In contrast, the signal has various representations depending on which sample structure $v_{\ell j}$ is chosen. As there is only a limited number of $v_{\ell j}$ that can be interpreted by humans, it is necessary to select them to verify the validity of the extracted signal. Thus, our definition of a signal contains a subjective step in the choice of an interpretable $v_{\ell j}$. However, the $v_{\ell j}$s themselves are generated in a fully data-driven manner.\par
The signal spectrum $u_{\ell i}$ extracted by the proposed method is not always a signal understandable to humans. Therefore, the validity of the extracted spectrum $u_{\ell i}$ is verified based on human-interpretable $v_{\ell j}$. The interpretable $v_{\ell j}$ is chosen subjectively. Note that subjectivity is introduced here to project the signal onto the human-interpretable components of the signal $v_{\ell j}$ while the extraction of $u_{\ell i}$ by the proposed method is completely objective and based solely on data. Even if the procedure selects on the basis $v_{\ell j}$ before selecting the spectrum $u_{\ell i}$, the same verification can be performed, so the basis $v_{\ell j}$ is selected first in the actual analysis procedure.\par

\subsubsection{Empirical procedure for detecting signals}
\label{sec_procedure}
\begin{algorithm}[H]
\caption{Procedure for detecting signals}
\label{alg1}
\begin{algorithmic}
\FOR{$M_{\mbox{res}}\leftarrow M$; $M_{\mbox{res}}>0$; $M_{\mbox{res}}\leftarrow M_{\mbox{res}}-1$}
\WHILE{$s\leq N_s$}
\STATE Generate a subset $D_{M_{\mbox{res}}}:=\{x_{ij}\mid i\in[1,N],\ j\in J_{\mbox{res}}\}$ of size $M_{\mbox{res}}$.
\STATE Centre each observation in $D_{M_{\mbox{res}}}$ across its $N$ variables to zero mean and apply transposed SVD.
\STATE Select the component $\ell=\ell_c$ using the criteria in Secs.~\ref{select_v_GCM} and~\ref{select_v_gene}.
\STATE Construct the $1-P_i$ histogram for $u_{\ell_c i}$ and extract significant variable-side outliers.
\STATE $s\leftarrow s+1$.
\ENDWHILE
\ENDFOR
\end{algorithmic}
\end{algorithm}

\begin{figure}[htb]
    \centering
    \includegraphics[width=1.0\linewidth]{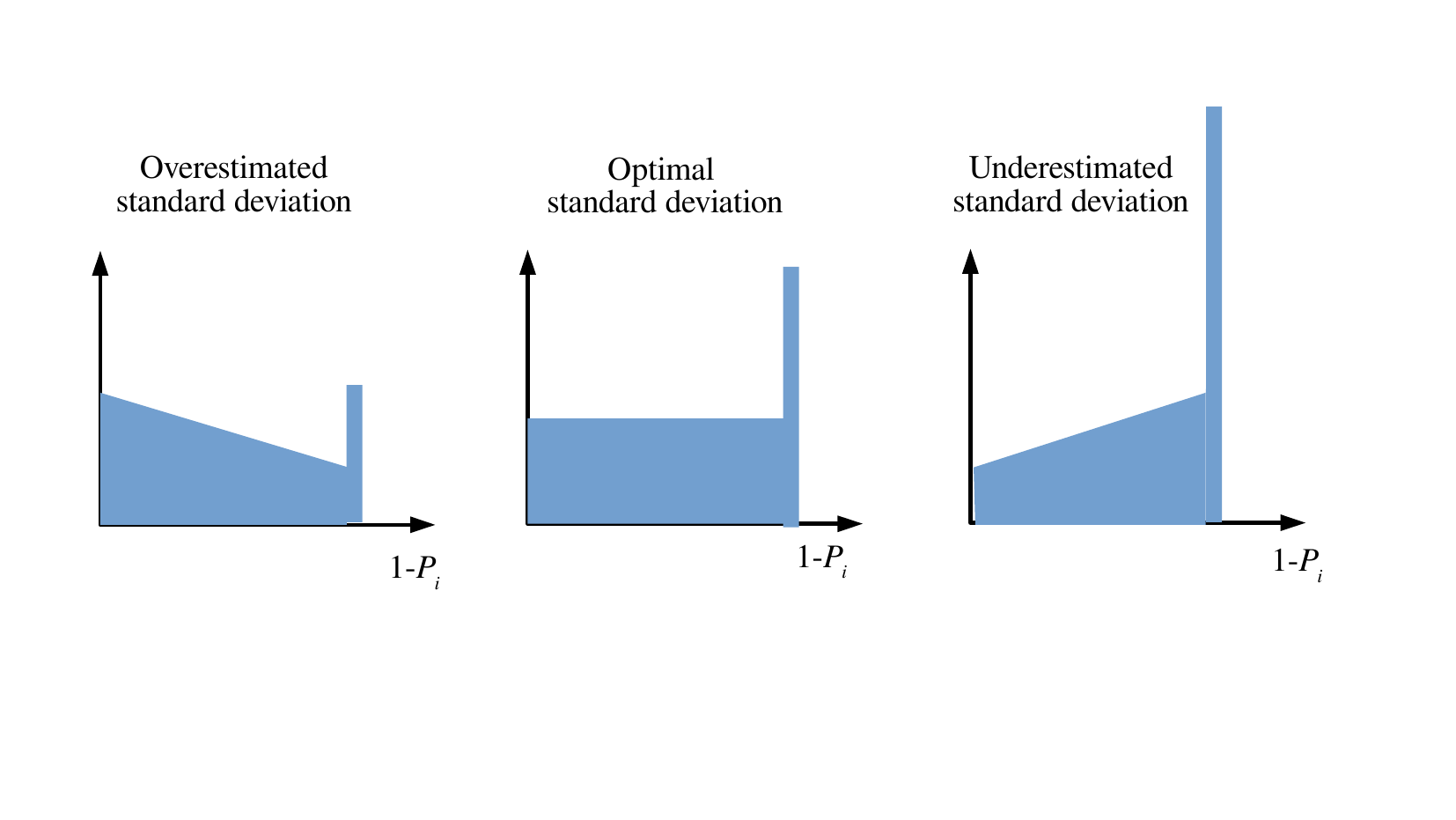}
    \caption{Dependence of the $1-P_i$ histogram on the estimated background SD.}
    \label{fig:hist}
\end{figure}
For each $M_{\rm res}>0$, subsets are drawn without duplication within a trial so that repeated observations do not create artificial correlations in the effective time-delay representation. The procedure is repeated over $N_s$ independently drawn subsets to estimate selection frequencies robustly with respect to the thinning pattern.\par

After subset construction, each observation is centred across its $N$ variables to zero mean, and transposed SVD is applied. This yields the variable-side set $D_{M_{\rm res}}^u:=\{u_{\ell i}\}_{i=1}^{N}$ obtained by projecting the $N$ points $\{(x_{i1},\ldots,x_{iM_{\rm res}})\}_{i=1}^{N}$ onto the $\ell$th principal direction. Outlier tests and BH correction are then applied to this distribution.\par
The value of $\sigma_\ell$ is optimised so that the non-outlier portion of the $1-P_i$ histogram is as flat as possible. As shown in Figure~\ref{fig:hist}, if $\sigma_\ell$ is overestimated, the histogram is biased toward small $1-P_i$; if it is underestimated, the histogram is biased toward large $1-P_i$. The optimisation procedure is detailed in Appendix~\ref{sec:sigma}.\par

After all trials are completed, the selection frequency of each variable is computed. The statistic $n_{>p\%}$ denotes the number of variables selected in more than $p\%$ of the trials. Its dependence on $M_{\rm res}$ is used to estimate the persistent component by extrapolation to $M_{\rm res}=0$. The extracted variable-side signals are interpreted through the corresponding sample-side vector $v_{\ell j}$. Implementations of the underlying feature-selection procedure are available in the Bioconductor packages TDbasedUFE and TDbasedUFEadv~\cite{10.3389/frai.2023.1237542,https://doi.org/10.18129/b9.bioc.tdbasedufe,https://doi.org/10.18129/b9.bioc.tdbasedufeadv}.

\section{Datasets for demonstration}
\label{sec_demo}
The RCS-GCM dataset provides a controlled test because the synchronised and chaotic variables are known from the generating dynamics. It is used here to compare the extrapolated non-vanishing component with the known synchronised variables. The TCGA pan-kidney dataset provides an empirical application with many variables and comparatively few samples. The RCS-GCM data retain their temporal ordering and provide known ground-truth dynamics and correlation structures. The TCGA profiles
are treated as temporally unordered samples from latent biological dynamics, since TCGA profiles apparently do not include time. Because permutations of the sample coordinates do not change the variable-side distribution or singular vectors, the same intrinsic-signal criterion is applied to both datasets. The RCS-GCM permits controlled validation, whereas the TCGA data provide an empirical test when the latent times and generating dynamics are not observed.
\subsection{RCS-GCM}
The GCM is a globally coupled system of multiple chaotic systems $x_{ij+1}=f(x_{ij},a)$~\cite{PhysRevLett.65.1391}, formulated as
\begin{eqnarray}
    x_{i j+1}   &= &  (1-g) f(x_{ij},a) + \frac{g}{N} \sum_{i'=1}^N  f(x_{i'j}, a), \\
    f(x,a) &= & 1-ax^2.
\end{eqnarray}
By adjusting the parameters $a$ and $g$, the GCM generates both globally synchronised states with persistent correlations and chaotic states with short correlations. To obtain a heterogeneous set of variable-side dynamics, a randomised map and coupling parameters are introduced as follows:
\begin{eqnarray}
    x_{i j+1}   &= &  g_{ii} f(x_{ij},a_i) + \frac{1}{N} \sum_{i'=1}^N g_{ii'} f(x_{i'j}, a_{i'}), \\
    g_{ii'} & =& (1-c) \delta_{ii'} + c \epsilon_{ii'},   \\
    a_i &\sim& U(1.7,1.95), \\
    f(x,a) &= & 1-ax^2,
\end{eqnarray}
where $\epsilon_{ii'}$ are uniform random numbers on $[0,1]$.\par
This study used $c=0.04$ and $N=10^4$. For an uncoupled map, the range $a_i\in[1.7,1.95]$ lies in the chaotic regime ($a_i>1.48$). The weak randomised coupling ($c=0.04$) preserves predominantly chaotic behaviour while allowing several groups of maps to synchronise, producing a heterogeneous mixture of ordered and random states. 100 observations were generated after drawing the initial values $x_{i0}$ independently from the uniform distribution on $[0,1]$, giving $X\in\mathbb{R}^{10^4\times100}$. The RCS-GCM realisation analysed in this study was generated with random seed 20260822 on a MacBook Pro (14-inch, 2023; Apple M2 Max, 96 GB memory) running macOS 14.7.6.\par
The observations were retained in their original order. Reordering the columns does not change the variable-side singular vectors used for selection, apart from the corresponding permutation of the sample-side vectors.

\subsubsection{Selecting the $\ell$ basis in the RCS-GCM}
\label{select_v_GCM}
The signal-extraction procedure was applied to multiple SVD components. 
Among the components examined, $\ell=1$ was selected because significant signal candidates were detected for this component. 
The corresponding sample-side vector $v_{1j}$ exhibited three distinct levels corresponding to the three synchronised states of the RCS-GCM (see Sec.~\ref{subsec_results_GCM}).

\subsubsection{Concrete procedures for signal component extraction in RCS-GCM data analysis}
\label{sec:conc_GCM}
After selecting $\ell$ using the procedure outlined in Section~\ref{select_v_GCM}, $P$-values were assigned to the $i$th feature as expressed in Eq.~(\ref{eq:Pi}) and corrected using the BH criterion. Features associated with adjusted $P$-values that were less than the threshold of 0.1 were selected. After all resampling trials (10,000 times) were completed, the frequency with which each feature was selected was computed for all $N$ features. As outlined in Section~\ref{intrinsic_signal_extraction}, $\sigma_\ell$ must be adjusted so that $h_n$ becomes uniform (Appendix~\ref{sec:sigma}). 
For smaller $M_{\mbox{res}}$, a small fraction of trials yielded pathological solutions with anomalously small optimised $\sigma_1$. These trials were identified automatically by two-cluster $k$-means classification using $\log_{10}q$ and the fraction of selected variables, where $q=\widehat{\sigma}_1/{\rm sd}(u_1)$; the classifier was trained using the $M_{\mbox{res}}=6$--11 trials and the same decision rule was subsequently applied to all $M_{\mbox{res}}$. Selection frequencies were computed using the retained trials as the denominator.
\par
A feature was regarded as a signal candidate when its adjusted $P$-value was below $0.1$. The statistic $n_{>1\%}$ denotes the number of features selected in more than $1\%$ of the resampling trials. Its dependence on $M_{\rm res}$ was used to estimate the non-vanishing component of the detected signal statistics in the zero-sample limit.

\subsection{Genomics data}
The gene-expression dataset contained 1,020 RNA-expression profiles with 20,531 numerical gene-expression variables,
\begin{equation}
x_{ij} \in D_{\rm RNA}:= \mathbb{R}^{20531 \times 1020}.
\end{equation}
The pathologic-M classification consisted of the categories m0, m1, and mx for profiles with available classification information. Among the 1,020 profiles, 627, 114, and 223 profiles were labelled m0, m1, and mx, respectively, while 56 profiles had missing or other labels. The latter profiles were retained in the original dataset but were not included in the balanced resampling. The pathologic-M labels were used to construct the balanced resampled subsets and to select and interpret the sample-side component $v_{\ell j}$: the regression of $v_{\ell j}$ on the pathologic-M category was performed for each resampled subset, whereas the variable-side coefficients $u_{\ell i}$ and their statistical significance were evaluated without using the labels after $\ell$ had been selected. 
Genomic data retrieved from TCGA was used, following the application setting of the original feature-selection framework~\cite{Taguchi2024,Taguchi2023.02.26.530076}. Specifically, the RTCGA library was used to access the RNA-expression profiles and pathologic-M categories for the pan-kidney cohort (KIPAN). ``patient.stage\_event.tnm\_categories.pathologic\_categories.pathologi\_m'' from RTCGA.clinical was used as the classification label.
For gene expression, resampling of $M_{\rm res}$ samples was performed for each of the three pathologic-M categories. Thus the actual number of samples used in each SVD was $3M_{\rm res}$; for example, $M_{\rm res}=20$ corresponds to 60 samples in the SVD.

\begin{table}[!htb]
\caption{Pathologic-M-category frequencies among RNA-expression profiles
with available classification labels. Profiles without one of the three retained labels
were retained in the original dataset but were not used in the balanced resampling
or in the regression on the pathologic-M category.}
    \centering
    \begin{tabular}{c|ccc}
       Label & m0 & m1 & mx \\\hline 
Frequency & 627 & 114 & 223  
 \end{tabular}
    \label{table:label_genome}
\end{table}

\subsubsection{Selecting the $\ell$ basis in genomics data analysis}
\label{select_v_gene}
The SVD component used for variable-side scoring was selected based on its association with the pathologic-M category. After resampling, SVD was applied to $x_{ij} \in \mathbb{R}^{N \times 3M_{\mbox{res}}}$. Each sample-side vector $v_{\ell j}$ was then analysed using a regression with categorical variables (RCV), in which one-hot encodings of the pathologic-M categories were used as explanatory variables:
\begin{equation}
    v_{\ell j} = a_\ell + \sum_s b_{\ell s} \delta_{sj} ,
\end{equation}
where $\delta_{s j}$ takes 1 when the $j$th sample belongs to the $s$th category, and 0 otherwise. The above RCV was evaluated by the overall $F$-test corresponding to the categorical regression. The component $\ell$ was selected as the one with the smallest overall $F$-test $P$-value in the RCV analysis. These $F$-test $P$-values were used only as a criterion for selecting the SVD component most strongly associated with the pathologic-M categories and were not interpreted as multiplicity-corrected inferential $P$-values. For Figure~\ref{fig:boxplot_genome} only, a representative trial was selected from the trials in which $\ell=4$ was chosen by taking the trial whose selected-component $P$-value was closest to the median of that subset; this display choice was not used in the downstream selection-frequency analysis.\par

\subsubsection{Concrete procedure for extracting signal components in a genomics dataset}
\label{sec:conc_gene}
After selecting $\ell$ using the procedure described in Section~\ref{select_v_gene}, $P$-values were assigned to the $i$th feature, as expressed in Eq.~(\ref{eq:Pi}), and corrected using the BH criterion. Features associated with the adjusted $P$-values less than the threshold of 0.01 were selected. After all resampling trials (1,000 times) were completed, the frequency of a selected feature was computed for all $N$ features. The statistic $n_{>95\%}$ was defined as the number of features selected more than 950 times among 1,000 trials, using an adjusted-$P$ threshold of $0.01$. No resampling trials were omitted in the genomics analysis. 
At the limit of $M_{\rm res}=0$, if $n_{>95\%}$ does not become zero, it can be concluded that the variable contains signal information. Incidentally, in the case of RCS-GCM, $p=1\%$, and Gene $p=95\%$, the criteria were different; however, the existence of the signal could be detected regardless of the criteria. The criterion $p$ was selected so that $n_{>p\%}$ did not become zero when $M_{\rm res}=0$.

\section{Results and discussion}
\label{sec_results}

\subsection{RCS-GCM: Results of signal extraction}
\label{subsec_results_GCM}

\begin{figure}[!htb]
    \centering
    \includegraphics[width=0.5\linewidth]{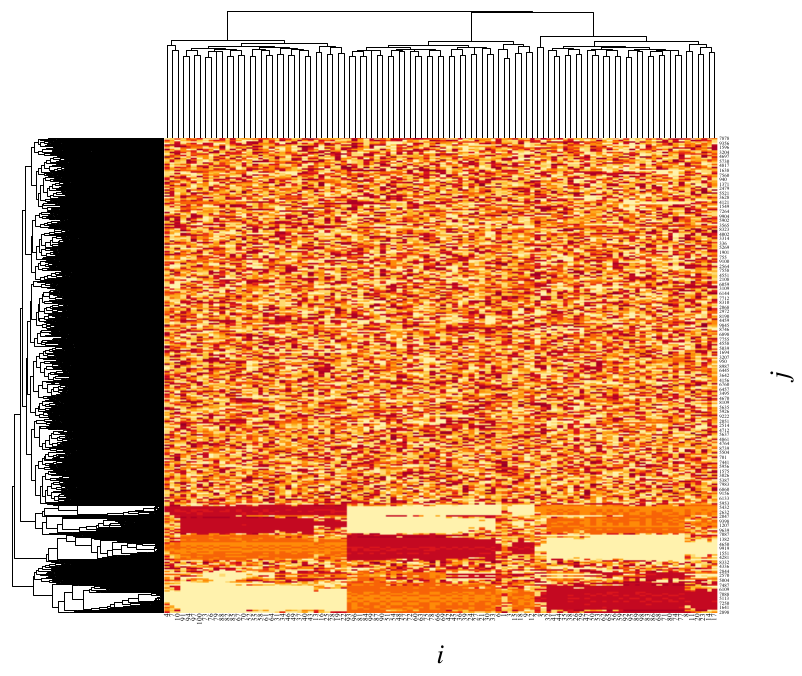}
    \caption{Heatmap of $x_{ij} \in \mathbb{R}^{10^4 \times 10^2}$ generated using the RCS-GCM.}
    \label{fig:GCM}
\end{figure}
Figure~\ref{fig:GCM} shows the heatmap of $x_{ij} \in\mathbb{R}^{10^4\times10^2}$ generated by the RCS-GCM. The rows were reordered by hierarchical clustering to place similar time series next to each another. The data contained many chaotic variables and a smaller set of synchronised three-state variables with long correlation times. The analysis tested whether these synchronised variables were selected as signals. As described in Section~\ref{select_v_GCM}, the sample-side vector $v_{1j}$ and the corresponding variable-side coefficients $u_{1i}$ were used for feature selection.
\par

\begin{figure}[htb]
    \centering
    \includegraphics[width=0.5\linewidth]{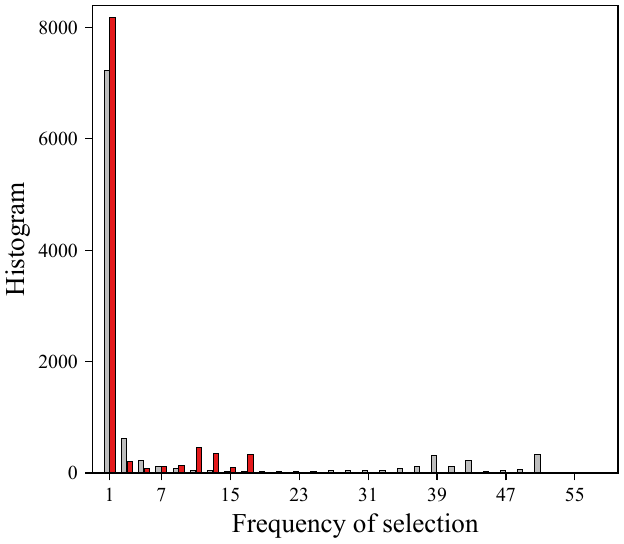}
    \caption{Histogram of the frequency with which individual features (RCS-GCM), $i$ are regarded as non-Gaussian. Red, $M_{\mbox{res}}=8$; Grey, $M_{\mbox{res}}=14$. The threshold $P$-value was 0.1, and the number of resampling trials was 10,000, although the frequency was scaled as a percentage.}. 
    \label{fig:FreqGCM}
\end{figure}
After all resampling trials were completed, the selection frequency was computed for all $N$ features. Figure~\ref{fig:FreqGCM} shows the frequency distribution for $M_{\rm res}=8$ and $14$. The number of frequently selected non-Gaussian features decreased as $M_{\rm res}$ was reduced, reflecting the loss of finite-correlation contributions at longer effective delays.\par
\begin{figure}[htb]
    \centering
    \includegraphics[width=0.4\linewidth]{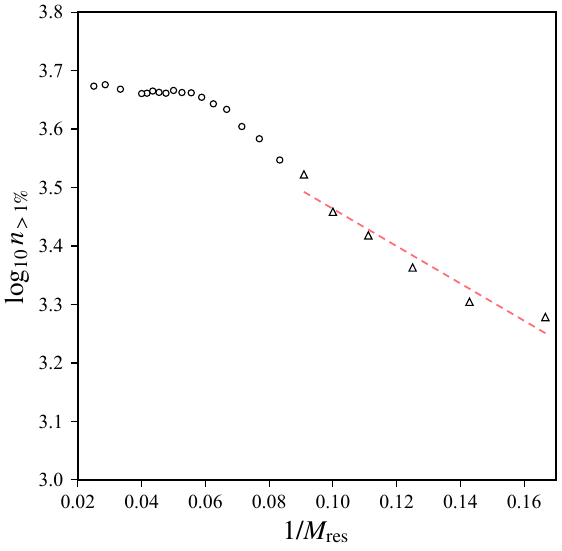}
    \caption{Number of features selected in more than $1\%$ of the trials as a function of $M_{\rm res}$ for the RCS-GCM ($M_{\rm res}=6,7,\ldots,25,30,35,40$). The vertical axis is $\log_{10}n_{>1\%}$ and the horizontal axis is $1/M_{\rm res}$. The red dashed line is the regression for $6\leq M_{\rm res}\leq11$ (black open triangles).}
    \label{fig:M_dep}
\end{figure}

The proposed small-sample analysis relies on components whose correlation lengths are long relative to the observation scale. The contribution of a component with a correlation length $\tau$ over an effective interval $T$ was modelled as
\begin{equation}
 \label{eq_def_GCMsignal}
    R(T;\tau)=\left\{\begin{array}{ll}
    1, & \tau=\infty,\\
    \alpha\exp(-T/\tau), & \tau<\infty,
    \end{array}\right.
\end{equation}
where $\alpha$ is a constant. This assumption is based on the empirical fact that many autocorrelations decay exponentially, and we numerically demonstrate below that this assumption is reasonable. As thinning observations gives $T\propto 1/M_{\rm res}$ (Figure~\ref{fig:time}),
\begin{equation}
\label{regress_P}
    R(M_{\rm res};\tau)=\left\{\begin{array}{ll}
    1, & \tau=\infty,\\
    \alpha\exp[-1/(M_{\rm res}\tau)], & \tau<\infty.
    \end{array}\right.
\end{equation}
Accordingly, the detected signal statistic can be represented as the sum of a non-decaying component and finite-correlation contributions,
\begin{eqnarray}
    n_{>\alpha\%} &\propto& \sum_i R(M_{\rm res};\tau_i)\\
    &=& n_0+\alpha\sum_{j:\tau_j<\infty}\exp\left[-\frac{1}{M_{\rm res}\tau_j}\right].
\end{eqnarray}
Near the small-$M_{\rm res}$ limit, the largest finite correlation length gives the leading decaying term,
\begin{eqnarray}
    n_{>\alpha\%}&\sim& n_0+\alpha\exp\left[-\frac{1}{M_{\rm res}\max(\tau_j)}\right],
    \label{eq:n_gcp}
\end{eqnarray}
where $n_0$ represents the non-vanishing component. Ignoring $n_0$ only for an initial assessment of the functional form gives
\begin{equation}
      \log_{10}n_{>1\%}\sim a-\frac{b}{M_{\rm res}},
      \label{eq:log}
\end{equation}
where $a$ and $b$ are constants. Figure~\ref{fig:M_dep} shows an approximately linear relation between $\log_{10}n_{>1\%}$ and $1/M_{\rm res}$ for $6\leq M_{\rm res}\leq11$, supporting the assumed finite-correlation decay. For larger $M_{\rm res}$, $n_{>1\%}$ approaches saturation because the total number of variables is finite. We therefore used $6\leq M_{\rm res}\leq11$ for the initial assessment of the asymptotic decay, before the saturation regime became dominant.\par
The threshold percentage $1\%$ and the regression range $6\leq M_{\rm res}\leq11$ were selected such that Eq.~(\ref{eq:n_gcp}) would be satisfied as much as possible. Consequently, the region where $M_{\rm res}>11$ (where $n_{>1\%}$ tends to take a constant value regardless of $M_{\rm res}$) was excluded from the regression region. This is because, as $M_{\rm res}$ increases, variables with shorter correlation lengths are selected as signals. Consequently, the premise underlying the outlier detection framework is that the majority follow a Gaussian distribution, whereas only a few variables exhibit deviant distributions. One might also wonder if $1\%$ is too small to regard features as non-Gaussian (i.e., signals). Nevertheless, because we need to decrease $M_{\rm res}$ to less than 10, the probability that individual features are selected as non-Gaussian decreases. Thus, a small probability was selected to consider results with sufficiently small $M_{\rm res}$ values.\par
The full fit includes the persistent component,
\begin{equation}
   n_{>1\%}=n_0+\alpha\exp\left[-\frac{\beta}{M_{\rm res}-M_{\rm lim}}\right],
   \label{eq:fit_GCM}
\end{equation}
where $M_{\rm lim}$ was fixed at zero in the present analysis and $\beta$ is related to the largest finite correlation length. The equivalent form used for non-linear regression is
\begin{eqnarray}
     \log(n_{>1\%}-n_0)&=&\log\alpha-\frac{\beta}{M_{\rm res}-M_{\rm lim}},\\
     M_{\rm res}&=&M_{\rm lim}-\frac{\beta}{\log(n_{>1\%}-n_0)-\log\alpha}.
     \label{eq:fit2_GCM}
\end{eqnarray}
\begin{figure}[!htb]
    \centering
    \includegraphics[width=0.4\linewidth]{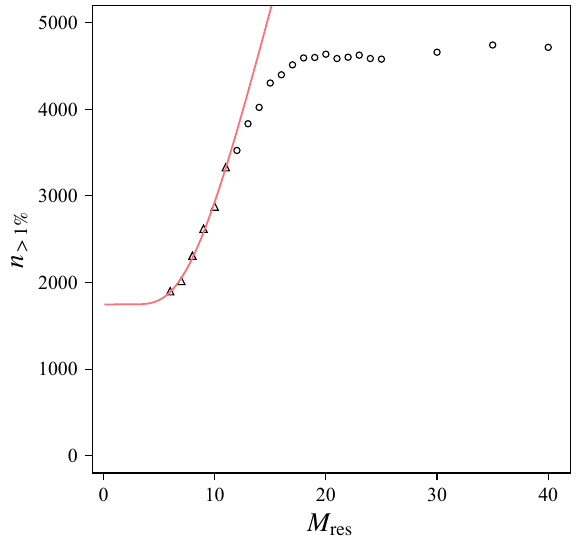}
    \caption{Fit for one realisation of the RCS-GCM dataset to Eq.~(\ref{eq:fit_GCM}). The regression was performed for $6\leq M_{\rm res}\leq11$ using Eq.~(\ref{eq:fit2_GCM}). For this realisation, the fitted limiting count was $n_0=1746.12$, with $\alpha=2.796\times10^4$, $\beta=31.64$, and $M_{\rm lim}=0$. }
    \label{fig:M_dep2}
\end{figure}
The non-linear fit for one realisation is shown in Figure~\ref{fig:M_dep2}. It yielded a positive fitted value $n_0=1746.12$, where $n_0$ represents the extrapolated number of variables remaining in the limit $M_{\rm res}\rightarrow0$. This contribution corresponds to the synchronised variables with long correlation times in the RCS-GCM. Models including $n_0$ also exhibited smaller Akaike Information Criterion (AIC) and Bayesian Information Criterion (BIC) values than models constrained to $n_0=0$ (AIC: $-0.61$ versus $11.35$; BIC: $-1.45$ versus $10.73$), when computed from the Gaussian residual likelihood of the inverse-$M_{\rm res}$ regression used for fitting. In this controlled example, the estimated non-zero limiting component is associated with the known synchronised variables, consistent with a component whose correlation length is effectively infinite relative to the observation scale.\par

In contrast, at the smallest analysed subsample size, $M_{\rm res}=6$, the variable-selection results from the 10,000-trial resampling analysis gave $n_{>1\%}(6)=1899$. Thus, 1746.12 is the extrapolated limiting count, whereas 1899 is the corresponding count directly obtained at finite $M_{\rm res}=6$. The closeness of these two values indicates that the result at $M_{\rm res}=6$ provides a reasonable finite-sample approximation to the estimated limiting component.\par

\begin{figure}[t]
    \centering
    \includegraphics[width=0.4\linewidth]{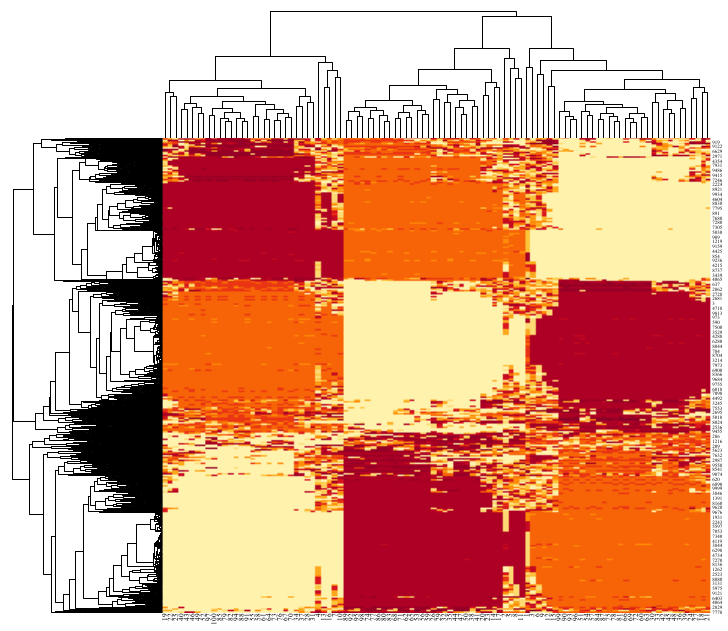}
    \caption{Heatmap of $x_{ij}$ for the 1899 variables counted as $n_{>1\%}$, ranked by their selection frequencies at $M_{\rm res}=6$. The selection frequencies were computed from the 10,000-trial resampling analysis as described above. The finite-$M_{\rm res}$ count $n_{>1\%}=1899$ is close to the extrapolated limiting count $n_0=1746.12$ obtained from the fit shown in Figure~\ref{fig:M_dep2}.}
    \label{fig:heatmap2}
\end{figure}

\begin{figure}[b]
    \centering
    \includegraphics[width=0.4\linewidth]{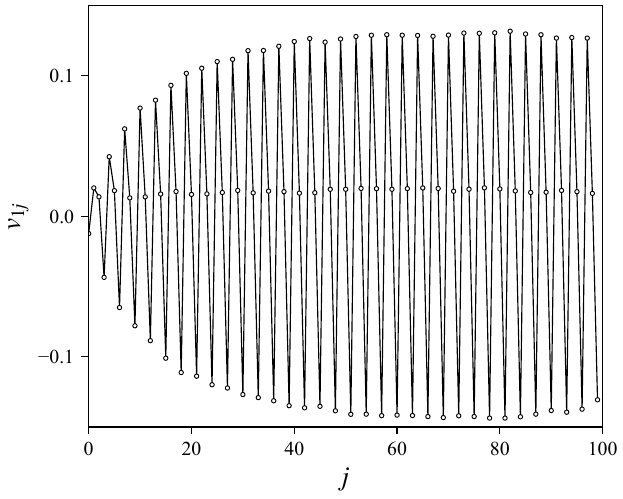}
    \caption{Sample-side vector $v_{1j}$ of the selected RCS-GCM component. The vector takes three distinct levels corresponding to the three synchronised states.}
    \label{fig:SVD}
\end{figure}
Figure~\ref{fig:heatmap2} shows the 1899 variables identified at $M_{\rm res}=6$. 
These variables are concentrated in the three synchronised states, indicating that the finite-$M_{\rm res}$ selection captures the long-correlation structures of the RCS-GCM. 
Figure~\ref{fig:SVD} shows the corresponding sample-side vector $v_{1j}$, which takes three distinct levels corresponding to these synchronised states. 
Thus, the three synchronised groups are represented by different levels of the same SVD component rather than by separate SVD components.\par

\subsection{Genomic data: Results of signal extraction}
\begin{figure}[b]
    \centering
    \includegraphics[width=0.5\linewidth]{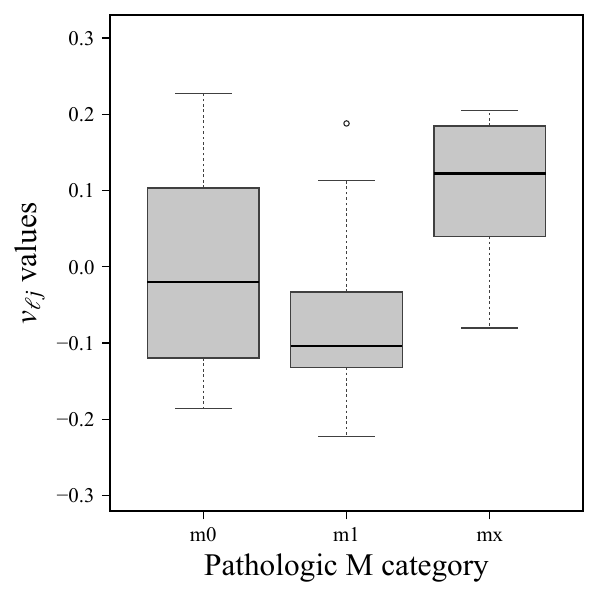}
    \caption{Representative boxplot of $v_{\ell j}$ for RNA-expression data at $M_{\rm res}=20$. In this resampling trial, the minimum-overall-$F$-test criterion selected $\ell=4$ ($P=4.56\times10^{-5}$). The displayed trial was chosen only for visualisation as described in Section~\ref{select_v_gene}.}
    \label{fig:boxplot_genome}
\end{figure}

Figure~\ref{fig:boxplot_genome} shows a representative resampling trial at $M_{\rm res}=20$. In this trial, the minimum-overall-$F$-test criterion selected component $\ell=4$ ($P=4.56\times10^{-5}$). The component was selected because its sample-side structure was associated with the pathologic-M labels (Table~\ref{table:label_genome}; Section~\ref{select_v_gene}). The labels were used for component selection and interpretation, whereas the variable-side coefficients $u_{\ell i}$ were scored from their distribution after $\ell$ was fixed. The selected SVD component was not the leading variance component, showing that the pathologic-M-associated structure was not a dominant variance component.\par

\begin{figure}[htb]
    \centering
    \includegraphics[width=0.5\linewidth]{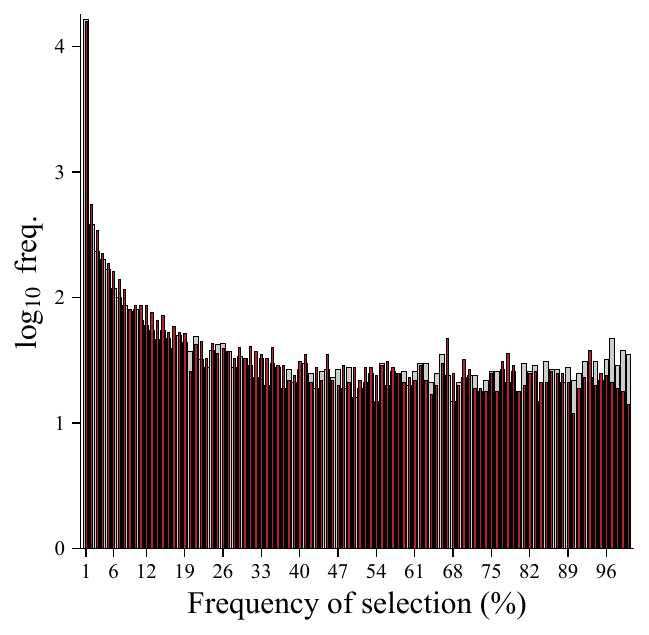}
    \caption{Logarithmic histogram of the frequency with which individual features (RNA gene expression),
    $i$, are regarded as non-Gaussian. Red, $M_{\mbox{res}}=20$; Grey, $M_{\mbox{res}}=40$. The threshold $P$-value was 0.01, and the number of trials (re-samplings) was 1,000; the selection frequency was expressed as a percentage.
}
    \label{fig:Freq1}
\end{figure}

The $P$-value for each variable was computed from Eq.~(\ref{eq:Pi}) using the optimised $\sigma_\ell$ (Appendix~\ref{sec:sigma}). Figure~\ref{fig:Freq1} shows the distribution of selection frequencies for $M_{\rm res}=20$ and $40$. Some variables were selected as non-Gaussian in nearly all resampling trials at $M_{\rm res}=20$, and five variables were selected in all resampling trials at $M_{\rm res}=40$; the total number of selected variables decreased as $M_{\rm res}$ decreased. Therefore, the gene-expression data contained variable-side deviations that remained detectable at small subsample sizes.\par

\begin{figure}[b]
    \centering
    \includegraphics[width=0.5\linewidth]{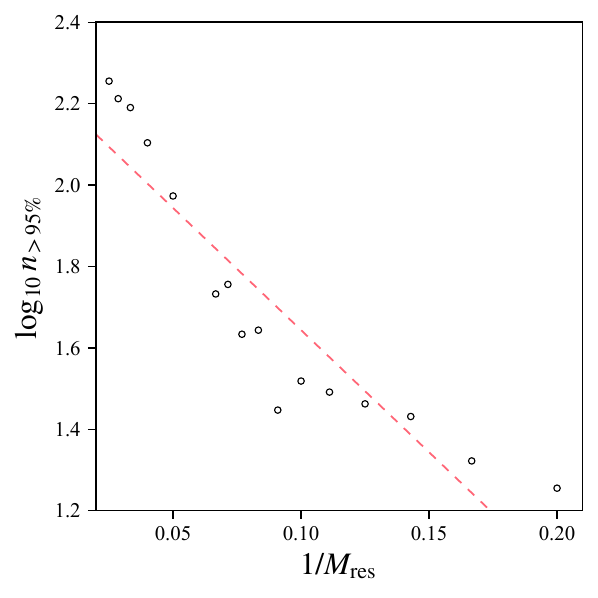}
    \caption{Dependence of $n_{>95\%}$ on $M_{\mbox{res}}$ for the RNA-expression data ($M_{\mbox{res}}=5,6,\dots,15,20,25,30,35,40$). Here, $n_{>95\%}$ denotes the number of variables selected in more than 95\% of the 1,000 resampling trials. The vertical axis is $\log_{10} n_{>95\%}$ and the horizontal axis is $1/M_{\mbox{res}}$. 
    The red broken line is the regression line assuming that $\log_{10} \left(n_{>95 \%} - N_A C\right) \sim \log_{10} \left(n_{>95 \%}\right)$ because the number of signal variables, $N_A$, was small. }
    \label{fig:gene_exp_dep}
\end{figure}

The temporal coordinates of the TCGA samples are not observed. However, the temporal ordering of the sample coordinates is not required in the present analysis, because a permutation of the columns of $X$ leaves $XX^{\mathsf T}$, and therefore the variable-side singular vectors, unchanged. We therefore treat the TCGA profiles as temporally unordered samples from latent biological dynamics and apply the same intrinsic-signal model as in the RCS-GCM analysis. Because the latent sampling intervals are not observed, the small-$M_{\rm res}$ asymptotic form for the TCGA data is derived from the Gaussian-background model rather than from explicitly known temporal intervals. Within the proposed model, the fitted constant $n_0$ represents a variable-side component that persists toward the small-sample limit and is consistent with an effectively infinite correlation length in the latent biological dynamics. This is a model-based inference rather than a direct measurement of the temporal correlation length. For the selected SVD component, the distribution of the variable-side coefficients $u_{\ell i}$ was modelled as a Gaussian background for noise variables together with an outlying distribution for signal variables [Figure~\ref{fig:randomwalk}]. 
Under this model, the number $n_{>x\%}$ of variables detected above the selection threshold is expressed as follows:
\begin{eqnarray}
    n_{>x \%}  &\sim& \sum_{j\in A}{C} + \sum_{i\in B}\int_{u^*}^\infty \frac{1}{\sqrt{2 \pi \sigma^2}}\exp\left(-\frac{1}{2\sigma^2}u_{\ell i}^2 \right) du_{\ell i},\\
   &=& N_A C + \frac{N_B}{2} {\rm erfc} \left(\frac{u^*}{\sqrt{2\sigma^2}}\right),
\end{eqnarray}
where $C$ is the appropriate constant, $u^*$ is the value of $u_{\ell i}$ corresponding to the threshold of the $P$-value for signal detection, A is the set of signal variables corresponding to the outlier distribution, and B is the set of noise variables corresponding to the Gaussian distribution with a mean value of zero. As the {\rm erfc} function becomes ${\rm erfc} (x) = \frac{e^{-x^2}}{x\sqrt{\pi}}\sum_{n=0}^\infty (-1)^n \frac{(2n)!}{n!(2x)^{2n}} \underset{x \gg 1}{\to} \frac{e^{-x^2}}{x\sqrt{\pi}}$ in the limit,
\begin{eqnarray}
   n_{>x \%}  &\sim& N_A C + \frac{N_B}{2} \frac{\exp\left[{-\left(\frac{u^*}{\sqrt{2\sigma^2}}\right)^2}\right]}{u^*\sqrt{\frac{\pi}{2\sigma^2}}}.
\end{eqnarray}
From this, the following equation is obtained:
\begin{eqnarray}
   \log \left(n_{>x \%} - N_A C\right)  \sim \log\left(\frac{N_B}{2}\right) -\left(\frac{u^*}{\sqrt{2\sigma^2}}\right)^2 - \log\left[u^*\sqrt{\frac{\pi}{2\sigma^2}}\right].
\end{eqnarray}
Here, $\mathcal{T}_{\rm rw}$ denotes the total elapsed time of the random walk associated with the $M_{\rm res}$ retained sample coordinates; assuming a constant time increment per coordinate, $\mathcal{T}_{\rm rw} \propto M_{\rm res}$. 
As the relationship $\sigma \propto \sqrt{\mathcal{T}_{\rm rw}}$ holds in random walks and $\mathcal{T}_{\rm rw} \propto M_{\rm res}$, $\sigma \propto \sqrt{M_{\rm res}}$ is obtained. Substituting this into the above equation, the following is obtained:
\begin{eqnarray}
   \log \left(n_{>x \%} - N_A C\right) &\sim& \log\left(\frac{N_B}{2}\right) -\left(\frac{u^*}{\sqrt{2kM_{\rm res}}}\right)^2 - \log\left[u^*\sqrt{\frac{\pi}{2kM_{\rm res}}}\right],\:k={\rm const.},\\
   &\underset{M_{\rm res} \ll 1}{\to}&  \log\left(\frac{N_B}{2}\right) -\left(\frac{u^*}{\sqrt{2kM_{\rm res}}}\right)^2,\\
   &=& a- \frac{b}{M_{\mbox{res}}},
\end{eqnarray}
where $a$ and $b$ are regression coefficients. For the gene-expression data, $n_{>95\%}$ was used, giving
\begin{equation}
    \label{eq_genefitting}
    \log_{10} \left(n_{>95 \%} - N_A C\right)   = a- \frac{b}{M_{\mbox{res}}},
\end{equation}
where $a$ and $b$ are optimised using linear regression~\cite{chambers:1992}. Figure~\ref{fig:gene_exp_dep} shows the dependence of $n_{>95\%}$ on $M_{\mbox{res}}$. The approximately linear relation between $\log_{10}n_{>95\%}$ and $1/M_{\mbox{res}}$ supports the finite-correlation decay assumed in the extrapolation model.\par

\begin{figure}[htb]
    \centering
    \includegraphics[width=0.5\linewidth]{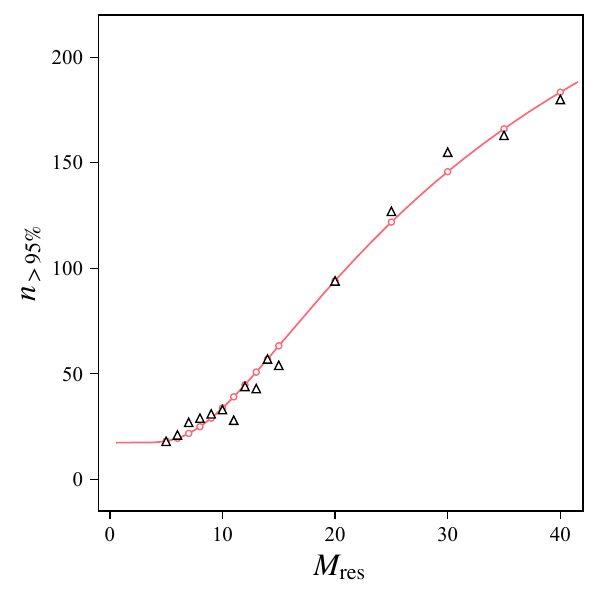}
    \caption{Non-linear fit of the data in Figure 12. The regression was performed using Eq.~(41), which is the rearranged form of Eq.~(39). The red curve and open circles show the fitted curve and corresponding fitted values, respectively. The fitted parameters were $n_0=17.36$, $\log_e \alpha=5.884$ ($\alpha=359.22$), $\beta=30.87$, and $M_{\rm lim}=0$.}
    \label{fig:gene_exp_dep2}
\end{figure}
To estimate the persistent component, we set $n_0:=N_A C$. Because direct fitting of Eq.~\eqref{eq_genefitting} with $n_0$ was numerically unstable, we instead used the following model: 
\begin{equation}
   n_{>95\%} = n_0 + \alpha \exp \left( - \frac{\beta}{M_{\mbox{res}} -M_{\mbox{lim}}} \right) \label{eq:fit_gene},
\end{equation}
where $M_{\mbox{lim}}$ was fixed at zero in the present analysis and $\beta$ simply replaces $b$ in Eq.~\eqref{eq_genefitting}. Eq.~\eqref{eq:fit_gene} is equivalent to:
\begin{eqnarray}
     \log \left ( n_{>95\%} - n_0 \right)  & =& \log \alpha  - \frac{\beta}{M_{\mbox{res}} -M_{\mbox{lim}}},   \\
   M_{\mbox{res}}&=&  M_{\mbox{lim}} -  \frac{\beta}{ \log \left ( n_{>95\%} - n_0 \right)  - \log \alpha}.  \label{eq:fit_gene2}
\end{eqnarray}
Figure~\ref{fig:gene_exp_dep2} shows the non-linear fit~\cite{bates1988nonlinear} to Eq.~(\ref{eq:fit_gene2}). Including $n_0$ reduced the AIC from 65.37 to 57.36 and the BIC from 67.69 to 60.45, with a fitted value $n_0=17.36$, when the information criteria were computed from the Gaussian residual likelihood of the inverse-$M_{\rm res}$ regression used for fitting. Therefore, the observed $M_{\rm res}$ dependence is better described by a model containing a non-vanishing term than by the model constrained to $n_0=0$. Within the proposed model, the fitted positive term $n_0=17.36$ indicates a variable-side component that persists toward the small-sample limit and is consistent with an effectively infinite correlation length in the latent biological dynamics. Because the sampling times are not observed, this is a model-based inference rather than a direct measurement of the temporal correlation length.\par

\section{Summary}
\label{sec_summary}
The variables of an $N\times M$ dataset were represented as $N$ points in an $M$-dimensional sample-coordinate space and transposed SVD was applied to their distribution. Signal candidates were defined as variable-side coefficients that departed from an estimated Gaussian background. Repeating the procedure over decreasing subsample sizes produced a selection-count curve $n_{>p\%}(M_{\rm res})$, from which a non-vanishing term $n_0$ was estimated.\par
For the RCS-GCM, the variables contributing to this term were concentrated in the known synchronised groups. For the TCGA pan-kidney data, the fit including $n_0$ was favoured by both AIC and BIC and gave $n_0=17.36$. This result shows that part of the variable-side non-Gaussian structure remained detectable under severe subsampling. In the RCS-GCM, this persistence was associated with known long-correlation dynamics. Both datasets were analysed under the common premise that their observations are sampled from underlying multivariate dynamics.
For the RCS-GCM, the temporal coordinates and long-correlation variables were known, allowing controlled validation. For the TCGA data, the temporal coordinates were unobserved, but their ordering was not required because the variable-side analysis is invariant to permutations of the sample coordinates. Within this model, the positive limiting term in the TCGA data is consistent with a long-correlation component in the latent biological dynamics.\par
Although gene expression profiles do not have explicitly defined time, they should be generated by a specific dynamic system (possibly the so-called gene regulatory network). If so, our framework offers advantages in detecting signals (ordered states), particularly when there are only a small number of samples, because the small sample size limit may correspond to a long time limit (Figure~\ref{fig:time}), where only the truly ordered state with an infinitely long correlation time can persist. In this sense, gene expression profiles that coincide with certain classification labels correspond to ordered states generated by the underlying dynamic systems.\par
Although we successfully extracted intrinsic signals that appeared to correspond to long-term correlations, the reason such data structures are inherent in real data, such as gene expression data, remains unexplained. Elucidating the mechanisms underlying the existence of such structures will contribute to understanding these complex phenomena.\par
Our study may also contribute to research in AI for Science. Recent advances in data-driven science have promoted the extraction of essential structures from observational data and the direct construction of physical models~\cite{doi:10.1126/sciadv.aay2631,brunton2016discovering,schmidt2009distilling,NEURIPS2020_c9f2f917,PhysRevLett.126.180604}. In particular, the selection of relevant variables or candidate terms constitutes an important part of physical-model inference itself, and numerous studies have addressed the data-driven inference of governing equations, including partial differential equations~\cite{rudy2017data, brunton2016discovering,uemura2015variable}. However, datasets obtained from complex physical phenomena are often high-dimensional, whereas the number of available observations is limited. The proposed framework may be effective for physical-model inference under such high-dimensional, small-sample conditions and may thereby contribute to the further development of AI for Science~\cite{Brunton2019,NEURIPS2020_c9f2f917}.

\subsection*{Acknowledgments}
This study was supported by JSPS KAKENHI (22K13979, 23K28150, 24K22309, 24H00247, 25K00986, and 25H01470); JST; PRESTO (JPMJPR212A); JST CREST (JPMJCR2431); and NEDO (JPNP22100843-0).

\subsection*{Conflicts of interest}
The authors declare no conflicts of interest regarding the publication of this paper.

\subsection*{Code and data availability}
The code used to generate the RCS-GCM data and to reproduce the GCM and TCGA analyses and figures, together with the exact raw RCS-GCM matrix analysed in this study, will be deposited in a persistent public repository [repository DOI to be inserted before publication]. The SHA256 checksum of the raw RCS-GCM matrix is \texttt{b36fe3e3a46ff7aa253ebe6598c645e54098d30e80749a6a53e33c76b72b9976}. The TCGA source data are publicly available from the Bioconductor Repository (DOI: 10.18129/B9.bioc.RTCGA).

\subsection*{Author contributions}
Both authors contributed equally to all aspects of this study.

\subsection*{Ethical statement}
Not applicable. This study did not involve human participants or animal subjects.

\appendix

\section{Optimisation of the SD}
\label{sec:sigma}

\begin{enumerate}
    \item Set the initial $\sigma_\ell$.
    \item Compute $P_i$ using Eq.~(\ref{eq:Pi}).
    \item Compute the histogram of $h_n$, where $h_n$ is the number of $i$s that satisfy $\frac{n- 1}{N_h} < 1-P_i \leq  \frac{n}{N_h}$. $N_h$ is the number of bins. Typically, $N_h =100$.
    \item Compute the adjusted $P_i$ considering multiple comparison corrections (e.g., BH criterion). 
    \item Exclude the BH-rejected tail and retain the contiguous background portion of the $1-P_i$ histogram. In the numerical implementation, a histogram bin was included in the background portion when its midpoint was below the cutoff separating the putative Gaussian background from the BH-rejected tail. Denote the retained set of bins by $H$.
    \item Compute the sample SD of the retained histogram counts, $\sigma_h$, as 
    \begin{eqnarray}
        \sigma_h & = &\sqrt{\frac{1}{|H|-1} \sum_{n\in H}\left (h_n -  \langle h \rangle_H \right)^2 },\\
         \langle h \rangle_H & = & \frac{1}{|H|} \sum_{n\in H} h_n .
    \end{eqnarray}
    \item Find $\sigma_\ell$ with the smallest $\sigma_h$.
\end{enumerate}
For the RCS-GCM analysis, the background SD was parameterised as $\sigma_1=q\,{\rm sd}(u_1)$ and $q$ was optimised over $[0.1,2.0]$ using a bounded one-dimensional search with \texttt{xatol}$=10^{-6}$ and at most 60 iterations. For the TCGA analysis, $\sigma_\ell$ was optimised directly over $[10^{-5},0.05]$ with \texttt{xatol}$=10^{-8}$ and at most 100 iterations. The final runs produced no optimizer failures and no solutions classified as lying at either search boundary.\par

\begin{figure}[tb]
    \centering
    \includegraphics[width=0.5\linewidth]{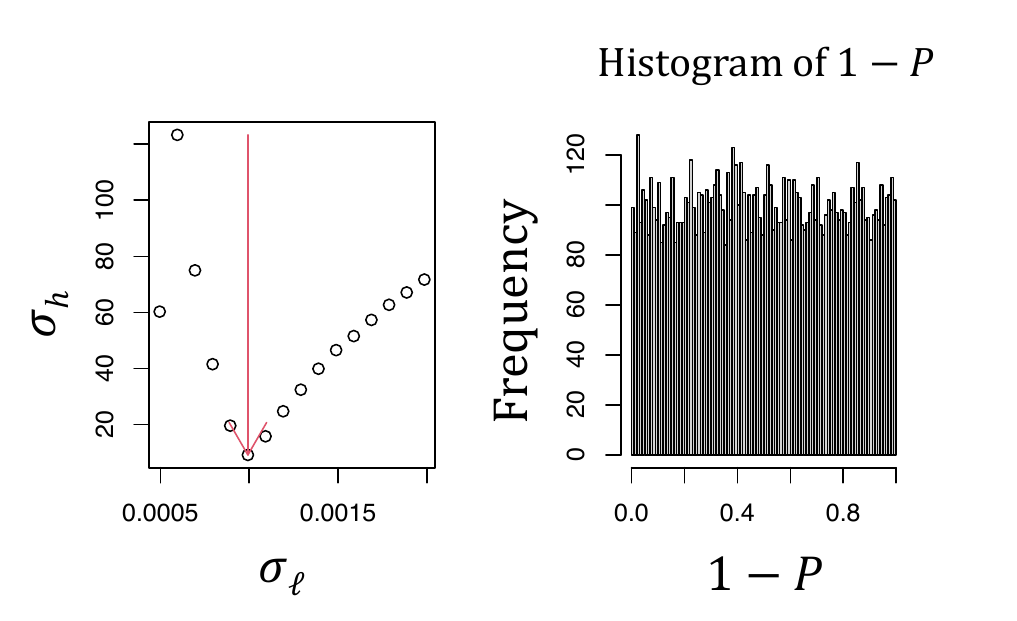}
    \caption{Demonstration of the optimisation of $\sigma_h$ for a pure Gaussian distribution. Left: dependence of $\sigma_h$ on $\sigma_\ell$. Right: histogram of $1-P_i$ with optimised $\sigma_\ell$. }
    \label{fig:opt}
\end{figure}

Figure~\ref{fig:opt} demonstrates the above optimisation procedure applied to 10,000 random features that are Gaussian with a zero mean and an SD ($\sigma_\ell$) of $1 \times 10^{-3}$. As expected, $\sigma_\ell = 1 \times 10^{-3}$ is associated with the minimum $\sigma_h$.

\section{Regression analysis}

The regression analyses of Eqs.~(\ref{eq:fit2_GCM}) and (\ref{eq:fit_gene2}) were performed in the equivalent inverse form
\begin{equation}
    M_{\mbox{res}} \sim   M_{\mbox{lim}} -  \frac{\beta}{ \log \left (n_{>p\%} - n_0 \right)  - \log \alpha}, 
\end{equation}
that is, $M_{\mbox{res}}$ was treated as the dependent variable and $n_{>p\%}$ as the independent variable. In the present implementation, $M_{\mbox{lim}}$ was fixed at zero, whereas $n_0$, $\log\alpha$, and $\beta$ were regarded as regression coefficients. The constraint $0\leq n_0<\min(n_{>p\%})$ was imposed so that $\log \left (n_{>p\%} - n_0 \right)$ remained well defined. The coefficients were estimated by multi-start Nelder--Mead minimisation of the squared residuals in $M_{\mbox{res}}$. AIC and BIC were computed assuming independent Gaussian residuals in this inverse-$M_{\mbox{res}}$ regression, including the residual-variance parameter in the parameter count; the model with $n_0$ has three regression coefficients and the model constrained to $n_0=0$ has two.

\section{Uniformity of the histogram of $P$-values under the null hypothesis~\cite{doi:10.1073/pnas.1530509100}}
\label{sec:hist}
Here, we briefly explain why the histogram of the $P$ -values computed using Eq.~(\ref{eq:Pi}) must be flat, as can be seen in Fig.~\ref{fig:opt}, if the null hypothesis that $u_{\ell i}$ is Gaussian is correct. This is not a field-specific truth but a consequence of the probabilistic nature of events.\par
\begin{figure}[htb]
    \centering
    \includegraphics[width=0.5\linewidth]{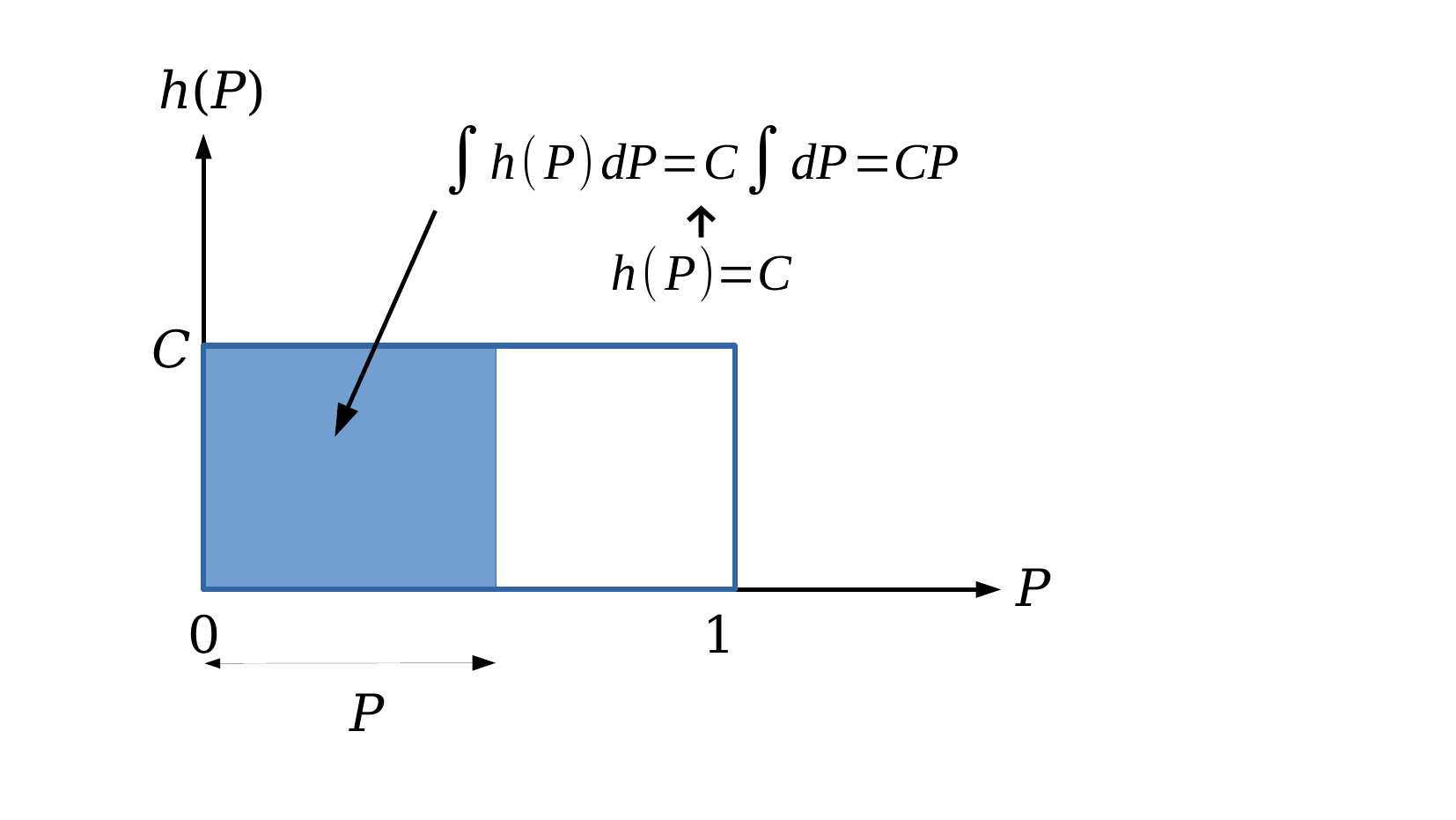}
    \caption{Under the null hypothesis, the probability integral transform implies that the $P$-values are uniformly distributed on $[0,1]$. Therefore, the expected histogram is constant, and the shaded area represents the cumulative probability up to a given $P$-value.}
    \label{fig:hist2}
\end{figure}

Suppose a certain random variable $x$ follows the distribution of $P(x)$, but we do not know $P(x)$ and incorrectly assume that $x$ follows $Q(x) \neq P(x)$. We then assign $P$-values to $x$ using $Q(x)$. Because of the identity
\begin{equation}
\int dP =P,
\end{equation}
the histogram of $P$ should always be flat (see Figure~\ref{fig:hist2}). However, if we incorrectly compute the left-hand side by counting the number of $P$-values computed using $Q$, this generally does not hold; that is,
\begin{equation}
\int dQ  \neq Q .
\end{equation}
Thus, by determining whether the histogram of $P$-values is flat, we can evaluate whether $Q=P$ (i.e., the correctness of $Q$).

\bibliography{apssamp}

@ARTICLE{10.3389/frai.2023.1237542,

AUTHOR={Taguchi, Y-h.  and Turki, Turki },

TITLE={Application note: {TD}based{UFE} and {TD}based{UFE}adv: bioconductor packages to perform tensor decomposition based unsupervised feature extraction},

JOURNAL={Frontiers in Artificial Intelligence},

VOLUME={6},

YEAR={2023},

URL={https://www.frontiersin.org/journals/artificial-intelligence/articles/10.3389/frai.2023.1237542},

DOI={10.3389/frai.2023.1237542},

ISSN={2624-8212}}

@PREAMBLE{
 "\providecommand{\noopsort}[1]{}" 
 # "\providecommand{\singleletter}[1]{#1}%" 
}

@misc{https://doi.org/10.18129/b9.bioc.tdbasedufe,
  doi = {10.18129/B9.BIOC.TDBASEDUFE},
  howpublished = {https://doi.org/10.18129/B9.bioc.TDbasedUFE},
  url = {https://bioconductor.org/packages/TDbasedUFE},
  author = {{Y-h. Taguchi}},
  title = {{TD}based{UFE}},
  publisher = {Bioconductor},
  year = {2023}
}

@misc{https://doi.org/10.18129/b9.bioc.tdbasedufeadv,
  doi = {10.18129/B9.BIOC.TDBASEDUFEADV},
  howpublished = {https://doi.org/10.18129/B9.bioc.TDbasedUFEadv},
  url = {https://bioconductor.org/packages/TDbasedUFEadv},
  author = {{Y-h. Taguchi}},
  title = {{TD}based{UFE}adv},
  publisher = {Bioconductor},
  year = {2023}
}

@book{bates1988nonlinear, 
title={Nonlinear Regression Analysis and Its Applications}, 
author={Bates, D.M. and Watts, D.G.}, 
series={Wiley Series in Probability and Statistics}, 
year={1988}, 
publisher={Wiley} }

@article{
doi:10.1073/pnas.1530509100,
author = {John D. Storey  and Robert Tibshirani },
title = {Statistical significance for genomewide studies},
journal = {Proceedings of the National Academy of Sciences},
volume = {100},
number = {16},
pages = {9440-9445},
year = {2003},
doi = {10.1073/pnas.1530509100},
URL = {https://www.pnas.org/doi/abs/10.1073/pnas.1530509100},
eprint = {https://www.pnas.org/doi/pdf/10.1073/pnas.1530509100}}

@book{chambers:1992,
  author = {Chambers, J.M. and Hastie, T.},
  description = {Statistical models in S - John M. Chambers, Trevor Hastie - Google Livres},
  isbn = {9780534167646},
  lccn = {91017646},
  publisher = {Wadsworth \& Brooks/Cole Advanced Books \& Software},
  series = {Wadsworth \& Brooks/Cole computer science series},
  title = {Statistical models in S},
  url = {http://books.google.fr/books?id=uyfvAAAAMAAJ},
  year = 1992
}

@article{PhysRevLett.65.1391,
  title = {Globally coupled chaos violates the law of large numbers but not the central-limit theorem},
  author = {Kaneko, Kunihiko},
  journal = {Phys. Rev. Lett.},
  volume = {65},
  issue = {12},
  pages = {1391--1394},
  numpages = {0},
  year = {1990},
  month = {Sep},
  publisher = {American Physical Society},
  doi = {10.1103/PhysRevLett.65.1391},
  url = {https://link.aps.org/doi/10.1103/PhysRevLett.65.1391}
}

@article {Taguchi2023.02.26.530076,
AUTHOR = {Taguchi, Y-h and Turki, Turki},
TITLE = {Integrated Analysis of Gene Expression and Protein–Protein Interaction with Tensor Decomposition},
JOURNAL = {Mathematics},
VOLUME = {11},
YEAR = {2023},
NUMBER = {17},
ARTICLE-NUMBER = {3655},
URL = {https://www.mdpi.com/2227-7390/11/17/3655},
ISSN = {2227-7390},
DOI = {10.3390/math11173655}
}

@book{Taguchi2024,
  title = {Unsupervised Feature Extraction Applied to Bioinformatics: A PCA Based and TD Based Approach},
  ISBN = {9783031609817},
  ISSN = {2522-8498},
  journal = {Unsupervised and Semi-Supervised Learning},
  publisher = {Springer International Publishing},
  author = {Taguchi,  Y-h.},
  edition = "2.",
  year = {2024}
}

@article{schmidt2009distilling,
  title={Distilling free-form natural laws from experimental data},
  author={Schmidt, Michael and Lipson, Hod},
  journal={Science},
  volume={324},
  number={5923},
  pages={81--85},
  year={2009},
  publisher={American Association for the Advancement of Science}
}

@article{
doi:10.1126/sciadv.aay2631,
author = {Silviu-Marian Udrescu  and Max Tegmark },
title = {{AI} {F}eynman: {A} physics-inspired method for symbolic regression},
journal = {Science Advances},
volume = {6},
number = {16},
pages = {eaay2631},
year = {2020},
doi = {10.1126/sciadv.aay2631},
URL = {https://www.science.org/doi/abs/10.1126/sciadv.aay2631},
eprint = {https://www.science.org/doi/pdf/10.1126/sciadv.aay2631}}

@inproceedings{NEURIPS2020_c9f2f917,
 author = {Cranmer, Miles and Sanchez Gonzalez, Alvaro and Battaglia, Peter and Xu, Rui and Cranmer, Kyle and Spergel, David and Ho, Shirley},
 booktitle = {Advances in Neural Information Processing Systems},
 editor = {H. Larochelle and M. Ranzato and R. Hadsell and M.F. Balcan and H. Lin},
 pages = {17429--17442},
 publisher = {Curran Associates, Inc.},
 title = {Discovering Symbolic Models from Deep Learning with Inductive Biases},
 url = {https://proceedings.neurips.cc/paper_files/paper/2020/file/c9f2f917078bd2db12f23c3b413d9cba-Paper.pdf},
 volume = {33},
 year = {2020}
}

@article{PhysRevLett.126.180604,
  title = {Machine Learning Conservation Laws from Trajectories},
  author = {Liu, Ziming and Tegmark, Max},
  journal = {Phys. Rev. Lett.},
  volume = {126},
  issue = {18},
  pages = {180604},
  numpages = {6},
  year = {2021},
  month = {May},
  publisher = {American Physical Society},
  doi = {10.1103/PhysRevLett.126.180604},
  url = {https://link.aps.org/doi/10.1103/PhysRevLett.126.180604}
}

@book{Brunton2019,
  title = {Data-Driven Science and Engineering: Machine Learning,  Dynamical Systems,  and Control},
  ISBN = {9781108422093},
  url = {http://dx.doi.org/10.1017/9781108380690},
  DOI = {10.1017/9781108380690},
  publisher = {Cambridge University Press},
  author = {Brunton,  Steven L. and Kutz,  J. Nathan},
  year = {2019},
  month = jan 
}

@article{achiam2023gpt,
  title={{GPT}-4 technical report},
  author={Achiam, Josh and Adler, Steven and Agarwal, Sandhini and Ahmad, Lama and Akkaya, Ilge and Aleman, Florencia Leoni and Almeida, Diogo and Altenschmidt, Janko and Altman, Sam and Anadkat, Shyamal and others},
  journal={arXiv preprint arXiv:2303.08774},
  year={2023}
}

@article{ba2016layer,
  title={Layer normalization},
  author={Ba, Jimmy Lei and Kiros, Jamie Ryan and Hinton, Geoffrey E},
  journal={arXiv preprint arXiv:1607.06450},
  year={2016}
}

@article{brunton2016discovering,
  title={Discovering governing equations from data by sparse identification of nonlinear dynamical systems},
  author={Brunton, Steven L and Proctor, Joshua L and Kutz, J Nathan},
  journal={Proceedings of the national academy of sciences},
  volume={113},
  number={15},
  pages={3932--3937},
  year={2016},
  publisher={National Acad Sciences}
}

@article{uemura2015variable,
  title={Variable selection for modeling the absolute magnitude at maximum of Type Ia supernovae},
  author={Uemura, Makoto and Kawabata, Koji S and Ikeda, Shiro and Maeda, Keiichi},
  journal={Publications of the Astronomical Society of Japan},
  volume={67},
  number={3},
  pages={55},
  year={2015},
  publisher={Oxford University Press}
}

@inproceedings{takens2006detecting,
  title={Detecting strange attractors in turbulence},
  author={Takens, Floris},
  booktitle={Dynamical Systems and Turbulence, Warwick 1980: proceedings of a symposium held at the University of Warwick 1979/80},
  pages={366--381},
  year={2006},
  organization={Springer}
}

@article{taguchi2022adapted,
  title={Adapted tensor decomposition and PCA based unsupervised feature extraction select more biologically reasonable differentially expressed genes than conventional methods},
  author={Taguchi, Y-h and Turki, Turki},
  journal={Scientific Reports},
  volume={12},
  number={1},
  pages={17438},
  year={2022},
  publisher={Nature Publishing Group UK London}
}

@book{Jolliffe2002,
  title={Principal Component Analysis},
  author={Jolliffe, Ian T.},
  publisher={Springer},
  edition={2nd},
  year={2002}
}

@book{Jackson1991,
  title={A User's Guide to Principal Components},
  author={Jackson, J. Edward},
  publisher={Wiley},
  year={1991}
}

@article{Pearson1901,
  title={On lines and planes of closest fit to systems of points in space},
  author={Pearson, Karl},
  journal={Philosophical Magazine},
  volume={2},
  number={11},
  pages={559--572},
  year={1901}
}

@article{Luxburg2007,
  title={A tutorial on spectral clustering},
  author={von Luxburg, Ulrike},
  journal={Statistics and Computing},
  volume={17},
  number={4},
  pages={395--416},
  year={2007}
}

@article{YataAoshima2013,
  author  = {Yata, Kazuyoshi and Aoshima, Makoto},
  title   = {{PCA} consistency for the power spiked model in high-dimensional settings},
  journal = {Journal of Multivariate Analysis},
  volume  = {122},
  pages   = {334--354},
  year    = {2013},
  doi     = {10.1016/j.jmva.2013.08.003}
}

@article{AoshimaEtAl2018,
  author  = {Aoshima, Makoto and Shen, Dan and Shen, Haipeng and Yata, Kazuyoshi and Zhou, Yi-Hui and Marron, J. S.},
  title   = {A survey of high dimension low sample size asymptotics},
  journal = {Australian \& New Zealand Journal of Statistics},
  volume  = {60},
  number  = {1},
  pages   = {4--19},
  year    = {2018},
  doi     = {10.1111/anzs.12212}
}

@article{Kefi2014,
  author  = {K{\'e}fi, Sonia and Guttal, Vishwesha and Brock, William A. and Carpenter, Stephen R. and Ellison, Aaron M. and Livina, Valerie N. and Seekell, David A. and Scheffer, Marten and van Nes, Egbert H. and Dakos, Vasilis},
  title   = {Early Warning Signals of Ecological Transitions: Methods for Spatial Patterns},
  journal = {PLoS ONE},
  volume  = {9},
  number  = {3},
  pages   = {e92097},
  year    = {2014},
  doi     = {10.1371/journal.pone.0092097}
}

@article{sauer1991embedology,
  title={Embedology},
  author={Sauer, Tim and Yorke, James A and Casdagli, Martin},
  journal={Journal of statistical Physics},
  volume={65},
  number={3},
  pages={579--616},
  year={1991},
  publisher={Springer}
}

@article{rudy2017data,
  title={Data-driven discovery of partial differential equations},
  author={Rudy, Samuel H and Brunton, Steven L and Proctor, Joshua L and Kutz, J Nathan},
  journal={Science advances},
  volume={3},
  number={4},
  pages={e1602614},
  year={2017},
  publisher={American Association for the Advancement of Science}
}

@article{bommasani2021opportunities,
  title={On the opportunities and risks of foundation models},
  author={Bommasani, Rishi and Hudson, Drew A and Adeli, Ehsan and Altman, Russ and Arora, Simran and von Arx, Sydney and Bernstein, Michael S and Bohg, Jeannette and Bosselut, Antoine and Brunskill, Emma and others},
  journal={arXiv preprint arXiv:2108.07258},
  year={2021}
}

@article{benjamini1995controlling,
  title={Controlling the false discovery rate: a practical and powerful approach to multiple testing},
  author={Benjamini, Yoav and Hochberg, Yosef},
  journal={Journal of the Royal statistical society: series B (Methodological)},
  volume={57},
  number={1},
  pages={289--300},
  year={1995},
  publisher={Wiley Online Library}
}

@book{hastie2009elements,
  title={The elements of statistical learning: data mining, inference, and prediction},
  author={Hastie, Trevor and Tibshirani, Robert and Friedman, Jerome H and Friedman, Jerome H},
  volume={2},
  year={2009},
  publisher={Springer}
}

\end{document}